\documentclass[%
amsfonts, amsmath, amssymb,
aps, 
prl, 
twocolumn, 
10pt, 
preprintnumbers, 
floats, 
altaffillsymbol, 
final, 
reprint, 
balancelastpage, 
flushbottom, 
noraggedfooter, 
showpacs, 
nobibnotes, 
nofootinbib, 
longbibliography, 
superscriptaddress,
]{revtex4-2}
\usepackage{Preamble}
\usepackage{tikz}
\usepackage{multirow}
\graphicspath{{./figures/}{./overlap_integral_figs/}}

\newcommand{\pbilby}{\texttt{parallel-Bilby}}
\newcommand{\dynesty}{\texttt{Dynesty}}
\newcommand{\nrsur}{\texttt{NRSur7dq4}}
\newcommand{\nrsurrem}{\texttt{NRSur7dq4Remnant}}

\newcommand{\ie}{{\it i.e.}~}
\newcommand{\eg}{{\it e.g.}~}
\newcommand{\vs}{{\it vs.}~}

\newcommand{\ZZ}{\ensuremath{{\cal Z}}}
\newcommand{\LL}{\ensuremath{{\cal L}}}
\newcommand{\BB}{\ensuremath{{\cal B}}}
\newcommand{\II}{\ensuremath{{\cal I}}}
\def\<#1>{\mathinner{\langle#1\rangle}}

\usepackage{layouts}

\newcommand*{\ztfflare}{ZTF19abanrhr}
\newcommand*{\theAGN}{J124942.3+344929}

\begin{abstract}
Recoiling remnants of black-hole mergers in dense environments can produce bright electromagnetic (EM) counterparts to the gravitational-wave (GW) emission. Significance assessments of such GW-EM candidates are restricted to time and sky-localization consistency, omitting the physics governing the EM emission process. Different emission mechanisms, however, impose different observability constraints on the remnant black-hole recoil and spin, which are gravitational-wave observables. We present a statistical framework that includes such parameters. We assess the consistency of the GW190521-\ztfflare\ pair with two types of emission processes: a Blandford-Znajek jet closely aligned with the final spin axis 
and a diffusive isotropic flare. 
Assuming the sky-location of \ztfflare, we find these mechanisms to be respectively strongly and moderately disfavoured with log-evidences $\log_{10}{\cal I}_{\rm jet} = - 1.65$ and $\log_{10}{\cal I}_{\rm diff} = -0.075$.
Combining these with odds for a common sky-location $\Omega$ we obtain respective combined odds $\log_{10} {\cal{O}}_{\Omega,{\rm jet}}= -1.17$ and $\log_{10} {\cal{O}}_{\Omega,{\rm diff}}= +0.39$ for a true GW-EM coincidence as opposed to a random one. Our method leverages a previously unexplored evidence axis to assess GW–EM associations and constrain both the physics powering flare mechanisms and the properties of AGNs.
\end{abstract}

\begin{document}
\title{Kick \& spin: new probes for multi-messenger black-hole mergers in AGNs}

\author{Samson H. W. Leong}
    \email{samson.leong@link.cuhk.edu.hk}
    \affiliation{Department of Physics, The Chinese University of Hong Kong, Shatin, N.T., Hong Kong}

\author{Juan Calder\'on~Bustillo}
    \email{juan.calderon.bustillo@gmail.com}
    \affiliation{Instituto Galego de F\'{i}sica de Altas Enerx\'{i}as, Universidade de Santiago de Compostela, 15782 Santiago de Compostela, Galicia, Spain}
    \affiliation{Department of Physics, The Chinese University of Hong Kong, Shatin, N.T., Hong Kong}

\maketitle
\section{Introduction}

Binary black-hole (BBH) in dense environments can produce an electromagnetic (EM) counterpart to the gravitational-wave (GW) signal through a variety of proposed mechanisms. A particularly attractive scenario in the recent years is that of BBHs within the accretion disks of Active Galactic Nuclei (AGN)~\cite{Graham:21gZTF,Graham2022:GWTC3,Leong2024:AGNConstraint,Ubach2025:self-lensing,Zhu2025:AGNEvidence,He2025:231123_AGN}. In this case, the recoiling remnant black hole (BH)~\cite{Netzer2015,Padovani2017:Name}, can interact with the surrounding gas, producing a transient EM flare~\cite{McKernan2019:RamPressure,Cabrera2024:S230922g}. 
Several such event candidates have been reported by the Zwicky Transient Facility (ZTF)~\cite{Graham:21gZTF,Graham2022:GWTC3}. Among these, particular attention has been paid to the pair candidate GW190521~\cite{LVK2020:GW190521} - \ztfflare~\cite{Graham:21gZTF}. The first is a mysterious GW signal originally reported by the LIGO-Virgo-KAGRA (LVK) Collaboration as a quasi-circular BBH with component BHs within the pair-instability supernova (PISN) gap leading to an intermediate-mass remnant black hole (IMBH) and with mild evidence for orbital precession~\cite{LVK2020:GW190521,LVK2020:21g_Properties}. 
Subsequent analyses, however, have found this signal is also consistent with a broad range of interpretations ranging from an eccentric BBH to an exotic-star merger~\cite{Gayathri2020:21g_Ecc,Romero-Shaw2020:21g_ecc,Gamba2021:21g_dyn,Nitz2020:21g_IMBH,Palmese2021:GW190521_H0,Fishbach2020:21g_straddling,Tanikawa2020:21g_PopIII,Anagnostou2020:21g_repeat,Barrera2022:21g_ancestral,Clesse2020:21g_PBH,Palmese2020:21g_ultradwarf,CalderonBustillo2020:Proca,CalderonBustillo2022:ProcaObs,Luna2024}. 
The second is an optical EM flare observed by the ZTF~\cite{Graham:ZTFScience,Bellm:ZTFResults} about 34 days after the GW event, originating from the AGN \theAGN, and excluded as false-positive with high confidence.
This flare was located at the 78\% sky localisation contours of GW190521, hence being identified as a plausible counterpart candidate.

\begin{figure*}[ht]
    \centering
    \includegraphics[width=\linewidth]{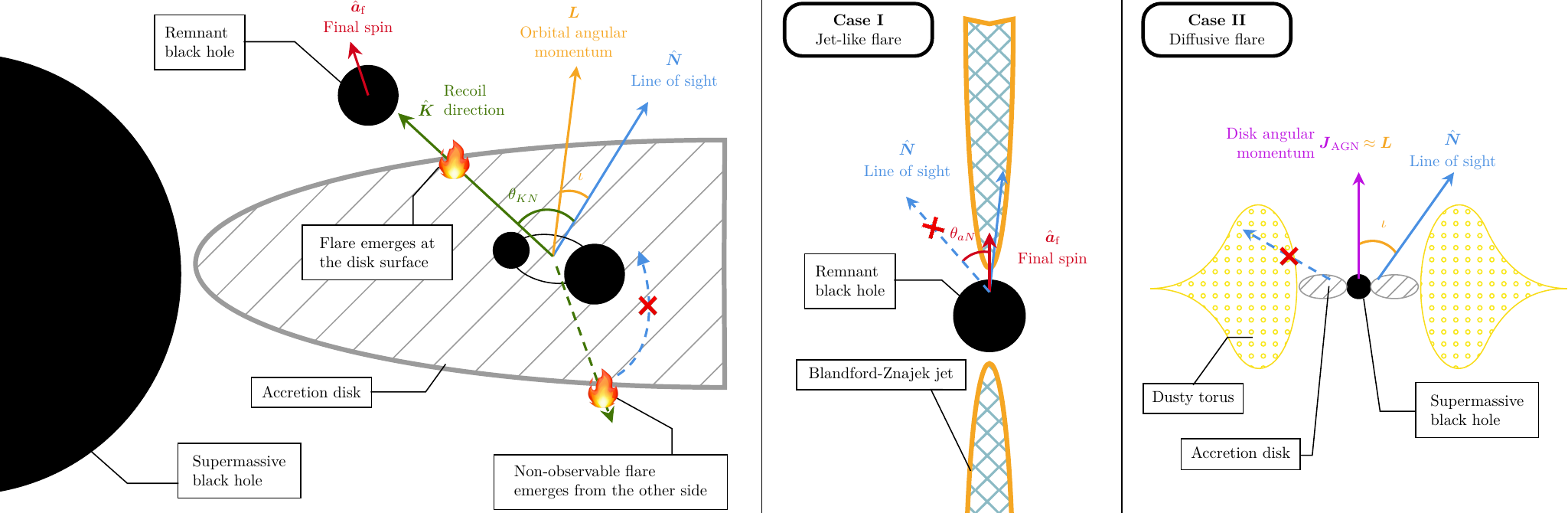}
    \caption{{\bf Illustration of the different AGN flare mechanisms and their relations to the remnant BH properties.} 
        Left: The remnant BH must be kicked out of the AGN disk on the side facing the observer. Center: For a jet-like flare, the observer line-of-sight must be within the jet cone, which is closely aligned with the remnant spin axis. Right: in the case of a diffusive isotropic flare, the observer must not be behind the torus surrounding the disk. 
    }
    \label{fig:illustration}
\end{figure*}

Accurate estimates of the statistical significance of such GW-EM pairs are crucial to determine the fraction of mergers occurring in AGNs~\cite{Cabrera2025:AGNConstraint}, probe accretion-disk physics~\cite{Ford2025:AGNGW,Vajpeyi2022:AGNProperties}, and provide independent cosmological measurements~\cite{Mukherjee2020:GW190521_H0,Chen2020:GW190521_H0,Gayathri2020:GW190521_H0,Palmese2021:GW190521_H0}. 

Currently, such estimates solely consider the consistency of the three-dimensional sky-location $\Omega$ of the GW and flare events~\cite{Ashton2020:Insufficient,Palmese2021:GW190521_H0}, completely omitting the physics underlying the emission mechanism and the flare visibility~\cite{Graham:21gZTF,Graham2022:GWTC3,Ahumada2024:O4a,Ashton2020:Insufficient,Veronesi2024:Causal,Cabrera2025:AGNConstraint}. Different emission mechanisms, however, impose specific constraints on the final BH properties, particularly its final spin and recoil direction, which can be directly extracted from GWs. This introduces a strong, but so far unexploited, connection between GW-derived remnant properties and the EM emission that can be used to a) characterise the emission mechanism b) further inform significance estimates.

Here, we introduce a formalism to include the mentioned remnant parameters and apply it to two types of emission mechanisms~\cite{Cabrera2024:S230922g}: a) a post-merger jet closely aligned with the final BH spin axis, powered by the Blandford-Znajek (BZ) process~\cite{Blandford1977:BZJet,Cruz-Osorio2021:M87Jet,Krolik2009:GRMHD,Mehlhaff2025,Shibata2025,Savard2025,Blandford2019:RelativisticJet} b) diffusive flares that leads to a rather isotropic emission. In both cases, the flare visibility imposes constraints on the recoil and spin of the remnant BH (Fig.~\ref{fig:illustration}). We test the consistency of the GW190521 parameters with these constraints. Assuming BZ jets display half-opening angle $\sim 10^\circ$ w.r.t.~the final spin \cite{Tchekhovskoy2008:jet,Tagawa:AGNSignature,Paschalidis2014:BHNS_jet}, we find this mechanism to be strongly rejected log Bayes' factor of $\log_{10} \II_{\rm jet} =-1.65$. For diffusive flares, assuming that the surrounding dusty torus requires the line-of-sight to form $<45^\circ$ with the AGN angular momentum, we obtain $\log_{10} \II_{\rm diff} =-0.87$. Combining this with the odds purely based on sky-location consistencies we obtain combined odds of $\log_{10} {\cal{O}}_{\Omega,{\rm jet}}= -1.17$ and $\log_{10} {\cal{O}}_{\Omega,{\rm diff}}= -0.39$. We repeat our analysis for different assumptions on the jet and AGN orientation. Finally, we demonstrate our method on simulated signals injected in zero noise, with signal-to-noise ratios of 15, 30 and 50, showing its strong power in removing false and enhancing true coincidences.

\section{Flare emission mechanisms and the AGN hypothesis~\label{sec:hypothesis}}
Asymmetric emission of linear momentum by BBHs in the form of GWs  imparts a gravitational recoil (or kick) to the BH remnant~\cite{Peres1962:RadiationRecoil,Thorne1980:multipole,Gonzalez:2006md,Lousto2007:recoil,Lousto2011:Hangup_kick,Lousto2012:AS_kick,Sperhake2010:superkick,Lousto2019:KickPhase}, whose magnitude and direction can be inferred through the GW signal~\cite{CalderonBustillo2018:tracing,Varma2020:Recoil,Varma2022:29m_Kick,CalderonBustillo2022:GW190412}. 
This can make the remnant BH escape the optically thick accretion disk of an AGN, leading to a detectable EM flare through various mechanisms including thermal emission from shocked disk gas, ram-pressure stripping of bound material, breakout radiation from Bondi–Hoyle accretion~\cite{Bartos2016:Rate,McKernan2019:RamPressure,Kimura2021:Breakout,Tagawa2024:JetBreakout}, and a BZ jet~\cite{Blandford1977:BZJet,Cruz-Osorio2021:M87Jet,Krolik2009:GRMHD,Mehlhaff2025,Shibata2025,Savard2025,Blandford2019:RelativisticJet}. We split these mechanisms into two types: (i) a post-merger BZ jet, collimated along the final spin axis with a characteristic opening angle (ii) processes that produce rather diffusive and isotropic flares. 

\subsection{Relation between flare visibility and remnant black-hole parameters}

The above choice of separation is motivated by the different constraints that the flare visibility imposes on the remnant parameters. In all cases, \textit{an observable flare requires the remnant to be kicked to observer-facing side of the optically thick disk}; as otherwise the emission is blocked by the optically-thick AGN disk, this translates into a threshold on the angle formed by the kick $\vb*K$ and the line-of-sight $\vb*N$ $\theta_{KN} \leqslant \theta_{KN}^{\rm th} \equiv \tan^{-1}(R_{\rm max}/h)$, where $h$ is the binary's height (from mid-plane) and $R_{\rm max}$ the maximal distance travelled before the flare onset. Under the thin-disk limit, this condition simplifies to $\theta_{KN}^{\rm th} \leq 90^\circ$.

Numerical simulations show that BZ jets display opening angles $\theta_{\rm jet,a} \simeq 10^{\circ}$~\cite{Tchekhovskoy2008:jet,Berger2013:grb,Tagawa:AGNSignature,Paschalidis2014:BHNS_jet}. A BZ jet is therefore visible only \textit{if the observer lies within the jet cone}, \ie~which requires $\overline{\theta_{aN}} \leqslant \theta_{aN}^{\rm th}$, where $\bar{\theta}=\min(\theta,\pi-\theta)$.
This condition does not apply to diffusive flares. However, these can be obscured by the torus surrounding the disk, which have typical half-opening  $\theta_{\rm AGN,torus}^{\rm th}\sim 30^\circ$–$60^\circ$~\cite{Zier2002:gr,Almeida2017:obscuration} with the angular momentum of the AGN $\vb*J_{\rm AGN}$.
Studies show that the BBH angular momentum $\vb*L$ can align with $\vb*J_{\rm AGN}$ -- resulting from viscous interactions, or Bardeen-Petterson effect~\cite{Bardeen1975,King2005:Aligning,McKernan2019:MCMC,Yang2019:HierarchicalAGN,McKernan2023:AGN_spin,Ford2025:AGNGW} -- the torus imposes two effective visibility conditions. 
First, $\iota \leq \iota^{\rm th} \simeq \theta_{\rm AGN,torus}^{\rm th}$, \textit{ensuring that the observer is not behind the torus}.
While we will explore wide ranges for these thresholds, for our reference results we will adopt $\theta_{KN}^{\rm th} = 90^\circ$ and $\iota^{\rm th}=45^\circ$, consistent with previous analyses~\cite{Graham2022:GWTC3}.
In summary, the above mechanisms impose the constraints:\\

{\it Case I.} Blandford-Znajek jet: 
\begin{equation}
    \theta_{KN} \leqslant \theta_{KN}^{\rm th} 
    \qqtext{and}
    \overline{\theta_{aN}} \leqslant \theta_{aN}^{\rm th}\ .
    \label{eq:caseI}
\end{equation}

{\it Case II a.} Diffusive flare:
\begin{equation}
    \theta_{KN} \leqslant \theta_{KN}^{\rm th}  + \overline{\iota}
    \qqtext{and}
    \overline{\iota} \leqslant \iota^{\rm th}\ .
    \label{eq:caseII}
\end{equation}

We note that the assumption $\vb*L \slantparallel \vb*J_{\rm AGN}$ requires misaligned spins for the kick to happen out-of-plane~\cite{Bruegmann2007:superkick,Lousto2011:Hangup_kick,Leong2025:KickVGW} and produce a visible flare, thus actually leading to a precessing $\vb*L$. While in this case it is possible that $\vb*L$ will remain roughly oriented with  $\vb*L \slantparallel \vb*J_{\rm AGN}$, we consider our reference (conservative) constraint based only on the kick direction: \\ 

{\it Case II b.} Diffusive flare (conservative):
\begin{equation}
    \theta_{KN} \leqslant \theta_{KN}^{\rm th}  \ .
\end{equation}

\section{Assessing coincidence odds: the overlap integral~\label{sec:overlap}}
We assess the chances of a true association under each mechanism -- as opposed to a random coincidence -- through Bayesian model selection. 
The association model imposes conditions on two sets of parameters. First,  the sky-location  coincidence imposes the ``sharp'' condition $\Omega$ must coincide with the ``sharp'' estimate $\Omega^{0}$ from \ztfflare~\cite{Ashton2020:Insufficient,Veronesi2024:Causal,Cabrera2025:AGNConstraint}. Second, the remnant parameters $\theta_{\rm rem}$ must be within the \textit{extended ranges} allowed by a specific EM mechanism. 
This distinction allows to factorise the evidence ratio for the combined set of parameters \vs a random coincidence as
\begin{equation}
    \BB^{\Omega,{\rm rem}}_{\rm BBH} \equiv \II_{\Omega,{\rm rem}} = \II_\Omega\,\II_{{\rm rem}\mid\Omega}\,
    \label{eq:full_overlap}
\end{equation}
where $\II_\Omega$ denotes the evidence for $\Omega = \Omega^{0}$ while $\II_{{\rm rem}\mid\Omega}$ denotes the evidence for the proposed mechanism hypothesis \textit{conditional on the AGN location}. This is given by the overlap integral:
\begin{equation}
    \II_\Omega\,\II_{{\rm rem}\mid\Omega} = \int \dd\theta_{\rm rem} \, \frac{p_{\rm GW}( \theta_{\rm rem} \mid d,\ \Omega^{\rm 0}) \, p_{\rm EM}(\theta_{\rm rem})}{\pi_{\rm GW}(\theta_{\rm rem})}\ .
    \label{eq:kick_overlap}
\end{equation} 
Above, $p_{\rm GW}(\theta_{\rm rem} | d,\Omega^{\rm 0})$ denotes the marginalised posterior distribution for the parameters $\theta_{\rm rem}$ conditional to the flare location $\Omega^{\rm 0}$ while $d$ denotes the GW data. Finally, $\pi_{\rm GW}(\theta_{\rm rem})$ denotes the prior distribution on $\theta_{\rm rem}$ induced by the priors on the masses, spins and source orientation  of an agnostic analysis while $p_{\rm EM}(\theta_{\rm rem})$ the corresponding prior induced by the emission-mechanism characteristic constraints on $\theta_{\rm rem}$ imposed by an observable flare. For details, please see Appendix I. 

We compute $\II_{\Omega}$ explicitly by performing two parameter inference  runs on GW190521: one allowing for agnostic priors on $\Omega$ and one fixing $\Omega = \Omega^{\rm 0}$, obtaining $p_{\rm GW}(\theta_{\rm rem} | d, \Omega^{\rm 0})$ from the latter. For details on these runs, please see Appendix II.

In the following, for each emission mechanism, we will respectively express $\II_{{\rm rem}|\Omega}$ and $\BB_{\rm BBH}^{\Omega,{\rm rem}}$ as $\II_{\rm jet, diff}$ and $\BB_{\rm jet, diff}$ with the corresponding $p_{\rm EM}$ given by Eqs.~\eqref{eq:caseI} and \eqref{eq:caseII}).
Finally, the relative probability for each emission mechanism is given by the ratio of $\II_{\rm jet} / \II_{\rm diff}$.

\section{GW190521 and \ztfflare}

\begin{figure}[t]
    \centering
    \includegraphics[width=\linewidth]{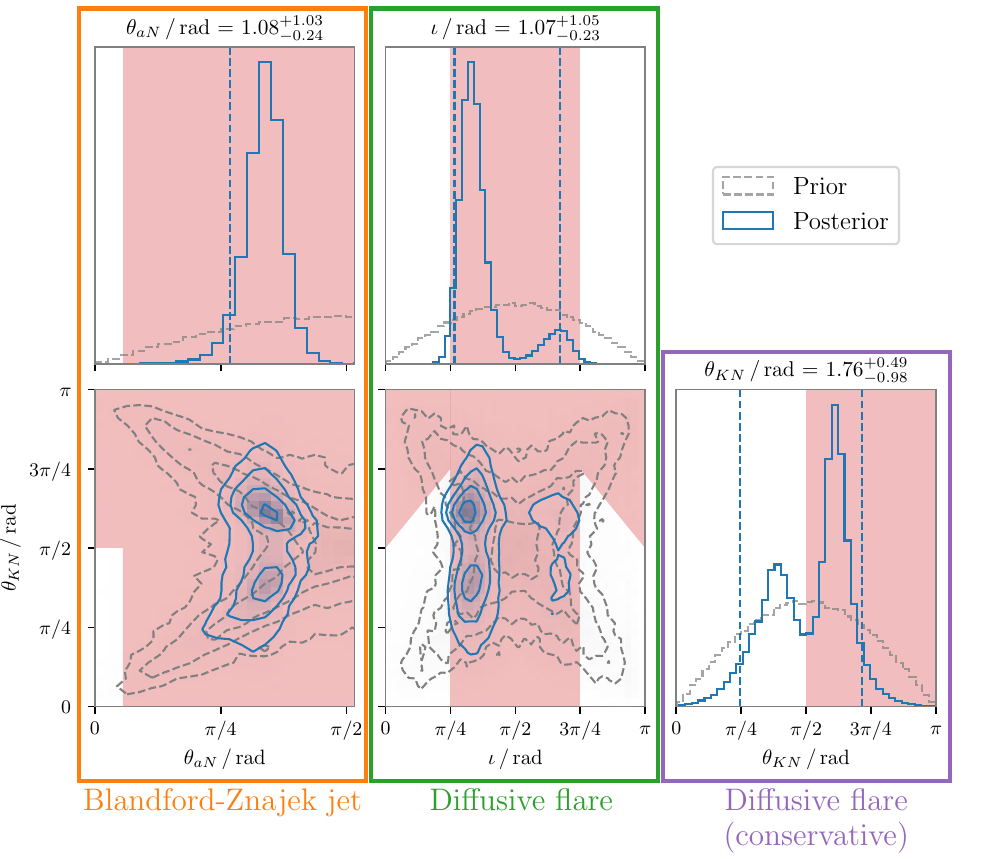}
    \caption{{\bf Constraints on the remnant parameters of GW190521  imposed by each emission mechanism.} 
        Two- and one-dimensional posterior distributions for the parameters $\theta_{aN},\ \iota$ and $\theta_{KN}$. The corresponding priors are shown in grey.
        Vertical dashed lines delimiting the symmetric 90\% credible intervals around the median, which are quoted at the top of the plot.  The two-dimensional contours denoting the $\{0.5, 1.0, 1.5, 2.0\}\,\sigma$ credible regions. 
        White areas denote the regions allowed the AGN hypothesis under jet and diffusive flare hypotheses, with forbidden regions represented in red.
        Note that show only half of the domain for $\theta_{aN}$ to ease the visualisation of the allowed region, and naturally the vertical line in the top-left panel delimits the left-credible interval. 
    }
    \label{fig:kick_posterior}
\end{figure}

Fig.~\ref{fig:kick_posterior} shows in blue the marginalised one and two-dimensional posterior distributions for the parameters $\theta_{aN}$, $\iota$ and $\theta_{KN}$ with priors in gray. 
Red/white shaded regions are excluded/allowed under each flare mechanism (Eqs.~\eqref{eq:caseI} \& \eqref{eq:caseII}). Note that measurements for $\theta_{aN}$ and $\iota$ are sufficiently informative that most white regions are rejected by the both mechanisms.

Table~\ref{tab:kick_overlap} shows the different relevant evidence ratios for each emission mechanism for GW190521. 
First, we find $\log_{10}\II_{\rm jet} = -1.649$, $\log_{10}\II_{\rm diff} = -0.865$ and $\log_{10}\II_{\rm diff, cons} = -0.075$. 
This is, assuming the AGN location, jet and diffusive flares are rejected strongly and moderately even under our non-conservative diffusive scenario.
Multiplying these by the evidence based on the sky-location consistency (first column), we obtain the combined evidences under each mechanism, shown in the last three columns. Under the BZ jet, any support coming from the sky-locations is completely erased, yielding $\log_{10}\BB_{\rm jet} = -0.06$. For the diffusive flare, we obtain a moderate support $\log_{10}\BB_{\rm diff} = 0.72$ and a rather strong support $\log_{10}\BB_{\rm diff} = 1.51$ under our conservative analysis. Finally, adding prior odds of $1/13$~\cite{Ashton2020:Insufficient}, we obtain posterior odds $\log_{10}{\cal{O}}_{\rm jet} \simeq -1.17$ rejecting the jet association\footnote{These prior odds account for the number of similar flares within the localisation volume of GW190521~\cite{Ashton2020:Insufficient,Graham:21gZTF}.}. For the diffusive case, we obtain mildly rejecting (inconclusive) evidences for our non-(conservative) analyses.
As a cross check, we repeat our analysis on the posterior samples obtained by Chen~{\it et\;al.}~\cite{Chen2020:GW190521_H0,Isi_21g_samples}, obtaining consistent results\footnote{In this case, we compute $\II_{\Omega}$ by comparing the evidences from the Chen~{\it et\;al.} analysis~\cite{Isi_21g_samples} (which has imposed $\Omega = \Omega^{0}$) and that from the LVK discovery paper~\cite{LVK2020:GW190521,GW190521:2020_data}.}. 

\begin{table*}[t]
    \centering
    \begin{tabular}{c||c || c | c | c || c | c | c}
        Samples & $\log_{10}{\II_{\rm \Omega_{3D}}}$ & $\log_{10}\II_{\rm jet}  $  & $\log_{10}\II_{\rm diff}$  & $\log_{10}\II_{\rm diff, cons}$  & $\log_{10}\BB_{\rm jet}$  & $\log_{10}\BB_{\rm diff}$ & $\log_{10}\BB_{\rm diff, cons}$ \\
        \toprule
        Ours  & 1.587 & -1.649 & -0.865 & -0.075   & -0.06 & 0.722 & 1.512 \\
        Chen+  & 1.632 & -1.471 & -0.942 & -0.267& -0.16 &  0.690 & 1.365
    \end{tabular}
    \caption{{\bf Assessing the GW190521-flare coincidence for different emission mechanisms.} We show the coincidence evidence based on sky-location (second column) and the evidence for jet and diffusive flares conditional on the flare location (third to fifth columns). The last three columns show the corresponding combined evidences. The columns with subscript ``diff,cons'' refer to the case where we only consider the kick direction for diffusive flares. The top and bottom rows show results obtained from our own analysis of GW190521 and from publicly available samples from Chen~{\it et\;al.}~\cite{Isi_21g_samples,GW190521:2020_data} respectively. Final odds are obtained by subtracting $\log_{10}(1/13) = -1.114$ from the last three columns.}
    \label{tab:kick_overlap}
\end{table*}

\section{Varying angle thresholds: parameter inference on the AGN and jet parameters~\label{sec:varying_logI}}

Both observations and simulations show that AGN disks their tori show that these can have diverse geometries. Moreover, while jet opening angles are known to display half-opening angles ${\cal{O}}(10)^\circ$, exact values can vary across different simulations. To account for this uncertainties, we recompute $\II_{\rm jet, diff}$ for varying threshold values. For the jet case, we consider $\theta^{\rm th}_{aN} \in [5^\circ, 40^\circ]$ and $\theta^{\rm th}_{KN} \in [50^\circ,90^\circ]$. Also, since different studies found the half-opening angle of the torus typically lies in $[30^\circ, 60^\circ]$, we set this as our range for $\theta^{\rm th}_{JN}$. 

\begin{figure}[ht]
    \centering
    \includegraphics[width=1.0\linewidth]{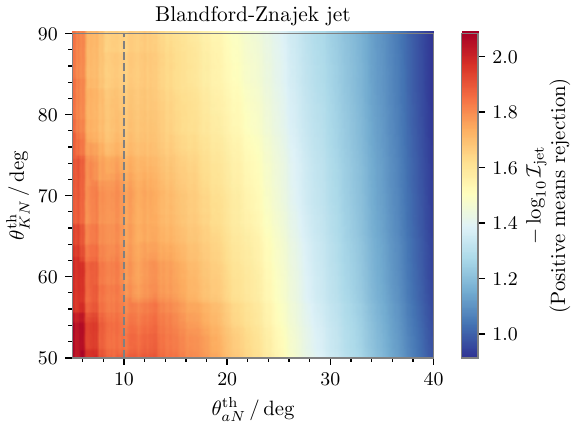}
    \includegraphics[width=1.0\linewidth]{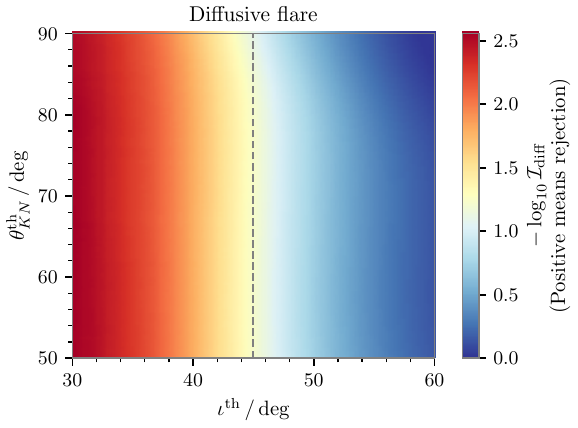}
    \caption{{\bf Evidence for a true association under flare mechanisms for varying remnant parameters thresholds}
        Inverse of the overlap integrals, in log scale $-\log_{10}\II$, as a function of varying thresholds on $\theta_{aN}$, $\iota$ and $\theta_{KN}$
        The top panel corresponds to the BZ jet (Eq.~\eqref{eq:caseI}) while the lower panel corresponds to the diffusive flare.
    }
    \label{fig:overlap_thresholds}
\end{figure}

Fig.~\ref{fig:overlap_thresholds} shows $\II_{\rm jet}(\theta_{aN}^{\rm th},\theta_{KN}^{\rm th})$ and $\II_{\rm diff}(\iota^{\rm th},\theta_{KN}^{\rm th})$ for GW190521. In both cases, the data is much more informative about $\iota$ and $\theta_{aN}$ than for $\theta_{KN}$. We see that consistency between GW190521 and the jet require require an \textit{unrealistically large jet opening angles}  $\theta_{{\rm jet},a}>40^{\circ}$. Under our reference conservative scenario, a diffusive flare would be allowed independently of $\theta_{KN}^{\rm th}$, while assuming orbit-AGN alignment would require either a ``wide AGN'' or ``short'' torus allowing for unobscured observers at $\iota>45^{\circ}$. Finally, note Fig.~\ref{fig:overlap_thresholds} does actually represent the (marginalised) likelihood for the threshold values $\LL(d|\theta^{\rm th}_{aN},\theta^{\rm th}_{KN})$, which are directly to the jet and AGN properties. Thus our method shall be interpreted as a way to performing indirect parameter inference on such properties, conditional to the association being true.

\section{Analysis of simulated signals}

We now reproduce our analysis on simulated BBH signals injected in zero-noise, using a three-detector Advanced LIGO-VIRGO network working at its sensitivity during GW190521. We randomly choose 6 BBH spin configurations. For these, we choose total mass consistent with GW190521 and mass-ratios $Q=1.5$ and $Q=3.0$ and true values of $\theta_{aN}^{\rm true} \in [0^\circ,30^\circ]$, yielding 48 distinct binary parameters, quoted in Table~\ref{tab:injection_params}. 
The values of $\theta_{aN}$ are intentionally chosen to be close to the limit imposed by the observability of the BZ jet. 
We focus on this mechanism as its test through $\theta_{aN}$ requires fewer assumptions on the disk and BBH geometry, making it more robust.  
Finally, we scale the luminosity distance of our injections to obtain signal-to-noise ratios (SNRs) of 15, 30 and 50, resulting in a total of 144 injections. For further details, please see Appendix II.\\

\begin{figure}[th]
    \centering
    \includegraphics[width=\linewidth]{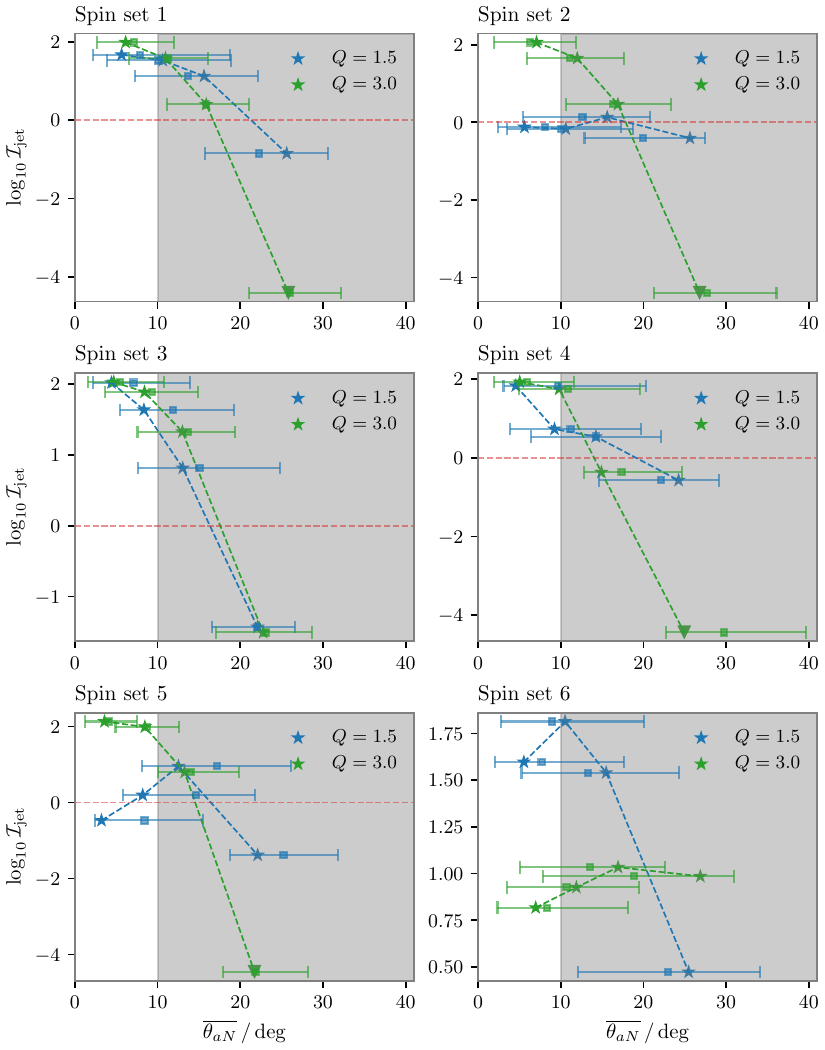}
    \caption{{\bf Identifying true Blandford-Znajek driven jets at high SNR.}
        Different panels correspond to BBHs with different spin configurations. 
        Blue (green) horizontal lines represent the symmetric $90\%$ credible intervals of $\overline\theta_{aN}$ obtained for simulated signals from BBHs with mass ratios $Q=1.5\ (3)$ injected in Advanced LIGO-Virgo.
        The median values are marked by the squares.
        For each injection, the overlap integral in support of the jet hypothesis is denoted by a star and plotted against the true value of $\theta_{aN}$. The white region $\theta_{aN} < 10^{\circ}$ corresponds to that allowed by the jet hypothesis.
    }
    \label{fig:injection_overlap}
\end{figure}

Fig.~\ref{fig:injection_overlap} shows $\II_{\rm jet}$ for our ${\rm SNR} = 50$ injections as a function of the true value of $\theta_{aN}$, denoted by stars\footnote{A few points are marked by an upside-down triangle to denote the upper bound we placed as an estimate of the overlap integral when there is insufficient posterior samples.}. Green (blue) horizontal lines respectively denote the symmetric posterior $90\%$ credible interval around the median for cases with $Q=1.5\ (3)$. We see that $\theta_{aN}$ come always unbiased\footnote{This is expected from~\cite{Leong2025:KickVGW,CalderonBustillo2022:GW190412}, which demonstrated accurate recovery of remnant parameters from NR simulations for ${\rm SNR} = 50$.}.

As expected, $\log_{10}\II_{\rm jet}$ decreases for growing $\theta^{\rm true}_{aN}$, crossing zero near the threshold value $\theta_{aN}=10^{\circ}$. Its value is largely determined by the width of the 90\% C.I.s and their intersection with the allowed region ($\theta_{aN} \leqslant 10^\circ$). For true jet-flares (\ie for $\theta_{aN}^{\rm true} < 10^{\circ}$) including remnant parameters can increase coincidence-odds by a factor up to $\log_{10}\mathcal{I}_{\rm jet} \approx +2$ w.r.t.~to those purely based on $\Omega$. This also implies that the existence of the jet mechanism can be established with strong evidence. Conversely, the jet-nature of the flare can be decisively ruled out with evidences factors $\log_{10}\mathcal{I}_{\rm jet} \lesssim -4$ for cases $\theta_{aN}^{\rm true} > 10^{\circ}$.

As expected, the broader parameter estimates for SNRs of 15 and 30 (see Appendix III) make $\II_{\rm jet}$ less conclusive for true cases and become strongly negative only for larger values of $\theta_{aN}^{\rm true}$. Nevertheless, true jet cases can be respectively supported with $\II_{\rm jet} \simeq 40 (100)$ while non-jet cases can be respectively penalised with $\II_{\rm jet} \simeq -3 (-100)$ within the limited range of angles shown in the figures. Decisive evidence against the jet hypothesis is always obtained for sufficiently large values of $\theta_{aN}^{\rm true}$.

\section{Discussion~\label{sec:discussion}}
EM counterparts to BBH signals in AGNs can be triggered by various mechanisms. Flare visibility imposes distinct constraints on the remnant BH parameters that depend on the particular mechanism that are currently ignored. We have introduced a new framework that incorporates remnant spin and recoil properties into GW–EM association studies. This complements existing 3D sky-localization methods, adding an independent physical filter based on the visibility of plausible EM emission mechanisms. Conversely, this framework can be used to confirm / rule-out the existence of certain mechanisms and constrain their parameters. Our method shall also be used to select between multiple candidate counterparts to GW signals and vice-versa~\cite{Graham2022:GWTC3,Cabrera2025:AGNConstraint}.

For GW190521, we find that the jet scenario is strongly disfavoured unless the jet has an unrealistic half-opening angle $\theta_{aN} > 45^\circ$. For diffusive flares, constraints from the AGN torus that depend on rather strong assumptions on the relative BBH-AGN orientation lead to moderate disfavour, tightening the current sky-localization association. Under more conservative assumptions that rely only on the kick direction, we cannot rule out a diffusive flare. Our conclusions are almost entirely driven by the spin direction, which is better constrained than the kick's. Our studies on simulated signals, show that our method is effective in enhancing true/rejecting false jet-flare candidates.

We note that while our results assume GW190521 is a quasi-circular BBH, this event is also consistent with an eccentric system~\cite{Gayathri2020:21g_Ecc,Romero-Shaw2020:21g_ecc,Gamba2021:21g_dyn}, which may impact our conclusions.

Finally, we stress that in all considered cases, the flare detectability requires the BH to be kicked towards the observer side of the AGN disk.
Such systems however lead to weaker GW signals than those recoiling away, further hindering the observation of true pairs (\eg~Fig.~1 in~\cite{CalderonBustillo2018:tracing}).
Similarly, the fact that the final BH must recoil off the disk implies that the BBH should display orbital precession~\cite{Sperhake2010:superkick,Leong2025:KickVGW}.
Current searches, however, are limited to aligned-spin binaries~\cite{Usman2015:PyCBC,Sachdev2019:gstlal}, limiting our sensitivity to precessing BBHs~\cite{CalderonBustillo2016:Precession,Harry2016:search,Chandra2020:NR_IMBH,Chandra2022:IMBHSearch}.

\section{Acknowledgements}
We thank Gonzalo Morr\'{a}s, Milton Ruiz, Tamara Bogdanovic and Emily McKey for useful comments and discussions.
SHWL acknowledges support by grants from the Research Grants Council of Hong Kong (Project No.~CUHK~14304622 and 14307923), and the Direct Grant for Research from the Research Committee of The Chinese University of Hong Kong.
JCB is supported by the Ramon y Cajal Fellowship RYC2022-036203-I,
and the research grant PID2020-118635GB-I00 from the Spain-Ministerio de Ciencia e Innovaci\'{o}n.
JCB is also supported by the Grant ED431F 2025/04 of the Galician CONSELLERIA DE EDUCACION, CIENCIA, UNIVERSIDADES E FORMACION PROFESIONAL. We also acknowledge support from the European Horizon Europe staff exchange (SE) programme HORIZON-MSCA2021-SE-01 Grant No. NewFunFiCO-101086251.
IGFAE is supported by the Ayuda Maria de Maeztu CEX2023-001318-M funded by MICIU/AEI /10.13039/501100011033.

This research has made use of data or software obtained from the Gravitational Wave Open Science Center (gwosc.org), a service of the LIGO Scientific Collaboration, the Virgo Collaboration, and KAGRA. 
This material is based upon work supported by NSF's LIGO Laboratory which is a major facility fully funded by the National Science Foundation, as well as the Science and Technology Facilities Council (STFC) of the United Kingdom, the Max-Planck-Society (MPS), and the State of Niedersachsen/Germany for support of the construction of Advanced LIGO and construction and operation of the GEO600 detector.
Additional support for Advanced LIGO was provided by the Australian Research Council.
Virgo is funded, through the European Gravitational Observatory (EGO), by the French Centre National de Recherche Scientifique (CNRS), the Italian Istituto Nazionale di Fisica Nucleare (INFN) and the Dutch Nikhef, with contributions by institutions from Belgium, Germany, Greece, Hungary, Ireland, Japan, Monaco, Poland, Portugal, Spain.
KAGRA is supported by Ministry of Education, Culture, Sports, Science and Technology (MEXT), Japan Society for the Promotion of Science (JSPS) in Japan; National Research Foundation (NRF) and Ministry of Science and ICT (MSIT) in Korea; Academia Sinica (AS) and National Science and Technology Council (NSTC) in Taiwan.
The authors are grateful for computational resources provided by the LIGO Laboratory and supported by National Science Foundation Grants PHY-0757058 and PHY-0823459.
We acknowledge the use of IUCAA LDG cluster Sarathi for the computational/numerical work.
The authors also acknowledge the use of the Gwave cluster provided by the Institute for Computational and Data Sciences at Penn State University and supported by National Science Foundation Grants OAC-2346596, OAC-2201445, OAC-2103662, OAC-2018299, PHY-2110594.
We acknowledge the use of computing facilities supported by grants from the Croucher Innovation Award from the Croucher Foundation Hong Kong.
This manuscript has LIGO-DCC number P2500748.

\appendix

\section{Appendix I: Combining sky-location and remnant quantities for multimessenger coincidence candidates~\label{app:overlap}}

In this appendix, we provide a detailed derivation of the overlap integral introduced in Eq.~\eqref{eq:kick_overlap} and discuss its applications. The main aspect of our derivation is that it does \emph{not} require a bijective mapping between the kick and the binary’s intrinsic parameters, offering a more robust foundation for applications such as those in Ref.~\cite{Lorenzo-Medina2025:Proca}.  

We anticipate the following key conclusion: while the priors of the sky locations $\Omega$ and derived parameters (kick, spin) $k$ are separable, the likelihood is not. Nevertheless, their \emph{overlap integrals remain separable}, provided that the integral over $k$ is computed \emph{conditioned on the true sky location} $\Omega^0$.

\subsection{Definitions and Preliminaries}

Throughout this appendix, we denote:  

\begin{itemize}
    \item $\Omega = \{\alpha, \delta, z\}$ as the 3D sky location.  
    \item $\theta$ as the intrinsic binary parameters (masses and spins).  
    \item $k(\theta)$ as a derived quantity from $\theta$, \eg remnant spin or recoil velocity.  
\end{itemize}

For simplicity, we will suppress marginalization over the inclination, phase, polarization, and coalescence time parameters. Thus, all likelihoods and posteriors are implicitly \emph{marginalized} over these extrinsic parameters.

We define a surjective mapping $K : \theta \mapsto k$. Then, for any probability density function $f(\theta)$ (prior or posterior), the corresponding density of $k$ is
\begin{equation}
f(k) = \int_\Theta f(\theta)\, \ind{K^{-1}(k)}(\theta) \, \dd\theta \,,
\label{eq:kick_density_def}
\end{equation}
where $\Theta$ denotes the domain of $\theta$, $\ind{A}(\cdot)$ is the indicator function, and $K^{-1}(k)$ denotes the pre-image of $k$. 
Note that this definition follows directly from the property of the indicator function~\cite{Ash1999:Measure}.
This ensures that $f(k)$ is properly normalized:

\begin{equation}
\int_{K(\Theta)} f(k)\, \dd k = \int_\Theta f(\theta)\, \dd\theta = 1\,.
\end{equation}

\subsection{Overlap Integrals}

Since the kick is an intrinsic property of the binary, it is \emph{independent of the sky location}, $\Omega \independent k$. This allows us to split the derivation into two steps: that pertaining the sky location and that pertaining remnant parameters.

\subsubsection{Sky Location Overlap}

Following Ref.~\cite{Ashton2020:Insufficient}, the overlap integral for a perfectly localized EM counterpart is
\begin{equation}
\II_\Omega = \frac{p(\Omega = \Omega^0 \mid d, C)}{\pi(\Omega = \Omega \mid C)}\,,
\end{equation}
where $d$ denotes the GW data, and $C$ represents the analysis context. 
By Bayes’ theorem,
\begin{equation}
\II_\Omega = \frac{\LL(d \mid \Omega = \Omega^0, C)}{\ZZ_{\rm full}(d \mid C)} \,,
\end{equation}
with $\LL(d \mid \Omega = \Omega^0, C)$ the likelihood at fixed sky location and $\ZZ_{\rm full}(d \mid C)$ the usual evidence over all sky locations. Explicitly, they are
\begin{align}
\ZZ_\Omega &= \int_\Theta \LL(d \mid \theta, \Omega = \Omega^0)\, \pi(\theta) \, \dd\theta\,, \\
\ZZ_{\rm full} &= \int_\Theta \int_\Phi \LL(d \mid \theta, \Omega)\, \pi(\theta)\, \pi(\Omega) \, \dd\theta\, \dd\Omega\,,
\end{align}
then we have
\begin{equation}
\II_\Omega = \frac{\ZZ_\Omega}{\ZZ_{\rm full}}\,.
\end{equation}

\subsubsection{Overlap Including Remnant Parameters}

We now include derived parameters $k$ (recoil, spin, etc.) in the overlap integral. Following Ref.~\cite{Ashton2020:Insufficient}, we define
\begin{equation}
\II = \iint \frac{p(\Omega,k \mid d) \, p_{\rm EM}(\Omega) \, p_{\rm EM}(k)}{\pi_{\rm GW}(\Omega)\, \pi_{\rm GW}(k)} \, \dd\Omega \, \dd k\,,
\end{equation}
where $p_{\rm EM}(\cdot)$ denotes distributions inferred from EM observations, and $\pi_{\rm GW}(\cdot)$ are GW priors.

Applying the definition of $p(k | d)$ (Eq.~\eqref{eq:kick_density_def}), and assuming $p_{\rm EM}(\Omega) = \delta(\Omega - \Omega^0)$, then integrating over $\Omega$, we obtain
\begin{equation}
\II = \iint \frac{\ZZ_{\rm 3D}\, p(\theta \mid \Omega = \Omega^0, d)}{\ZZ_{\rm full}} \frac{p_{\rm EM}(k)}{\pi_{\rm GW}(k)} \ind{K^{-1}(k)}(\theta)\, \dd \theta\, \dd k \,.
\end{equation}
Integrating over $\theta$ yields the \emph{kick overlap integral}, $\II_k$:
\begin{equation}
{\cal I} = {\cal I}_{\rm 3D} \, {\cal I}_k = {\cal I}_{\rm 3D} \int p(k \mid \Omega = \Omega^0, d) \frac{p_{\rm EM}(k)}{\pi_{\rm GW}(k)} \, \dd k \,.
\label{eq:kick_overlap_def}
\end{equation}
This justifies our definition in the main text (Eq.~\eqref{eq:full_overlap}).

\subsection{Implementation with EM-Informed Priors}

In this work, the EM observation defines an \emph{implied prior} over the remnant parameters:
\begin{equation}
p_{\rm EM}(\theta_{\rm rem}) \propto \pi_{\rm GW}(\theta_{\rm rem}) \, \ind{\Theta_H}(\theta_{\rm rem})\,,
\end{equation}
where $\Theta_H$ is the domain defined by the emission mechanism (\eg jet or diffusive flare), Eqs.~\eqref{eq:caseI} \& \eqref{eq:caseII}. The normalization constant is
\begin{equation}
{\cal N}_H = \int_{\Theta_H} \pi_{\rm GW}(\theta_{\rm rem}) \, \dd \theta_{\rm rem}\,.
\end{equation}

The kick overlap integral (Eq.~\eqref{eq:kick_overlap_def}) then becomes
\begin{equation}
\II_H = \frac{1}{{\cal N}_H} \int_{\Theta_H} p(\theta_{\rm rem} \mid d, \Omega^0)\, \dd \theta_{\rm rem}\,.
\end{equation}

The two integrals above can be efficiently estimated using \emph{importance sampling}:
\begin{equation}
\II_H \approx \frac{N^\pi}{N^p} \frac{\sum_i \ind{\Theta_H}(\theta_{{\rm rem},i}^p)}{\sum_j \ind{\Theta_H}(\theta_{{\rm rem},j}^\pi)}\,,
\end{equation}
where $i$ runs over posterior samples $\theta_{{\rm rem},i}^p \sim p(\theta_{\rm rem} | d, \Omega^0)$ and $j$ over prior samples $\theta_{{\rm rem},j}^\pi \sim \pi_{\rm GW}(\theta_{\rm rem})$.

\begin{itemize}
    \item In cases with no posterior samples in $\Theta_H$, we approximate an upper bound as $\II_H \approx 1/N^p$. These are denoted as the upside-down triangles in Fig.~\ref{fig:injection_overlap}.
    \item This procedure generalizes naturally to multiple remnant parameters.
\end{itemize}

\subsection{Conclusions}

\begin{enumerate}
    \item The overlap integral factorizes into sky location and remnant contributions: $\II = \II_\Omega \, \II_{\rm rem}$. With the notation in the main text, this translates into Eq.~\eqref{eq:full_overlap}.
    \item Posterior distributions of derived quantities are well-defined using pre-images of the mapping $\theta \mapsto k$, even if the mapping is not bijective. This was indeed exploited to compute Bayesian evidences in \eg~\cite{CalderonBustillo2022:GW190412,CalderonBustillo2024:Mirror}.
    \item The method naturally incorporates EM-informed priors and can be implemented via importance sampling.
    \item Factorization over multiple remnant parameters is valid if approximate independence holds. We note this does not apply in our case. 
\end{enumerate}

This derivation provides a rigorous and general framework for combining GW posteriors with EM constraints on remnant parameters.

\section{Appendix II: Parameter Estimation Details~\label{app:pe}}
We analyse 8\,s of publicly available strain data around GW190521 sampled at 1024\,Hz with the \pbilby~software~\cite{bilby1,bilby2,pbilby}. 
We use the \nrsur~waveform model~\cite{NRSur7dq4}, which is directly fitted to numerical simulations of BBHs including orbital precession and higher-order modes with, with a starting frequency of 11\,Hz.
We extract the remnant parameters using the corresponding model \nrsurrem~\cite{NRSur7dq4}. 
We sample the parameter space using the \dynesty~sampler~\cite{dynesty} with 8192 live points. Our priors on all parameters follow those used by the LVK the GW190521 discovery paper~\cite{LVK2020:GW190521}. To obtain the posterior conditional on $\Omega = \Omega^{0}$ and the corresponding Bayesian evidence, we perform the same run with the corresponding parameters fixed to the location of \ztfflare, given by the right ascension, declination, and redshift $(\alpha,\delta,z) = (3.358,\ 0.608,\ 0.438)$.

\section{Appendix III: Further details on the analysis of simulated signals~\label{app:inj_setup}}
All of our injections are computed through the \texttt{NRSur7dq4} waveform model. 
The spin parameters are chosen randomly from the isotropic distribution, which are displayed in Table.~\ref{tab:injection_params}.
In addition, all injections share the same azimuthal angle, sky locations, and polarisation angle. All these fixed values (except for the azimuth) correspond to the maximum likelihood parameters for the event GW190521 from the analyses we performed in Ref.~\cite{CalderonBustillo2020:Proca,CalderonBustillo2021:21gFlare}.

In practice, we cannot directly set the value of $\theta_{aN}$ for our injections, as this is determined through the masses and spins of the BBH.
We overcome this by using $\theta_{JN}$ as a proxy, given by the good agreement between $\theta_{JN}$ right and $\theta_{aN}$ shown in the Appendix of Ref.~\cite{Leong2025:KickVGW}.
For each of the six spin configurations, we varied the observer's inclinations as $\theta_{JN} \in \{ 5^\circ,\ 10^\circ,\ 20^\circ,\ 25^\circ\}$. 

\setlength{\tabcolsep}{5pt}
\begin{table}[t]
    \centering
    \begin{tabular}{c|| c c |c c | c c}
        Spin set & $a_1$ & $\theta_1$ &  $a_2$ & $\theta_2$ &  $\phi_{12}$ & $\phi_{JL}$ \\
        \toprule
        1 & 0.278 & 1.818 & 0.287 & 0.834 & 2.486 & 4.924 \\
        2 & 0.277 & 1.449 & 0.242 & 1.240 & 2.915 & 0.353 \\
        3 & 0.983 & 1.467 & 0.035 & 0.905 & 2.103 & 1.191 \\
        4 & 0.329 & 1.037 & 0.188 & 1.130 & 1.851 & 1.242 \\
        5 & 0.891 & 1.093 & 0.385 & 0.578 & 0.458 & 1.345 \\
        6 & 0.468 & 2.473 & 0.349 & 0.834 & 3.473 & 0.975 \\
    \end{tabular}
    \caption{{\bf Spin parameters of the simulated signals used in our injection study}. 
        The term $a_{i}$ denotes the spin magnitude of the $i$-th black hole while $\theta_{i}$ denotes the angle the spin forms with the orbital angular momentum $\vb*{L}$ of the binary.
        The two last two columns, $\phi_{12}$ and $\phi_{JL}$, are respectively the angles between the two spins, and between the total and orbital angular momenta, projected onto the binary orbital plane.
        All parameters are quoted at a reference frequency of 11\,Hz.
    }
    \label{tab:injection_params}
\end{table}

\subsection{Results on simulated signals with SNRs 15 and 30}
\begin{figure*}
    \centering
    \includegraphics[width=.48 \linewidth]{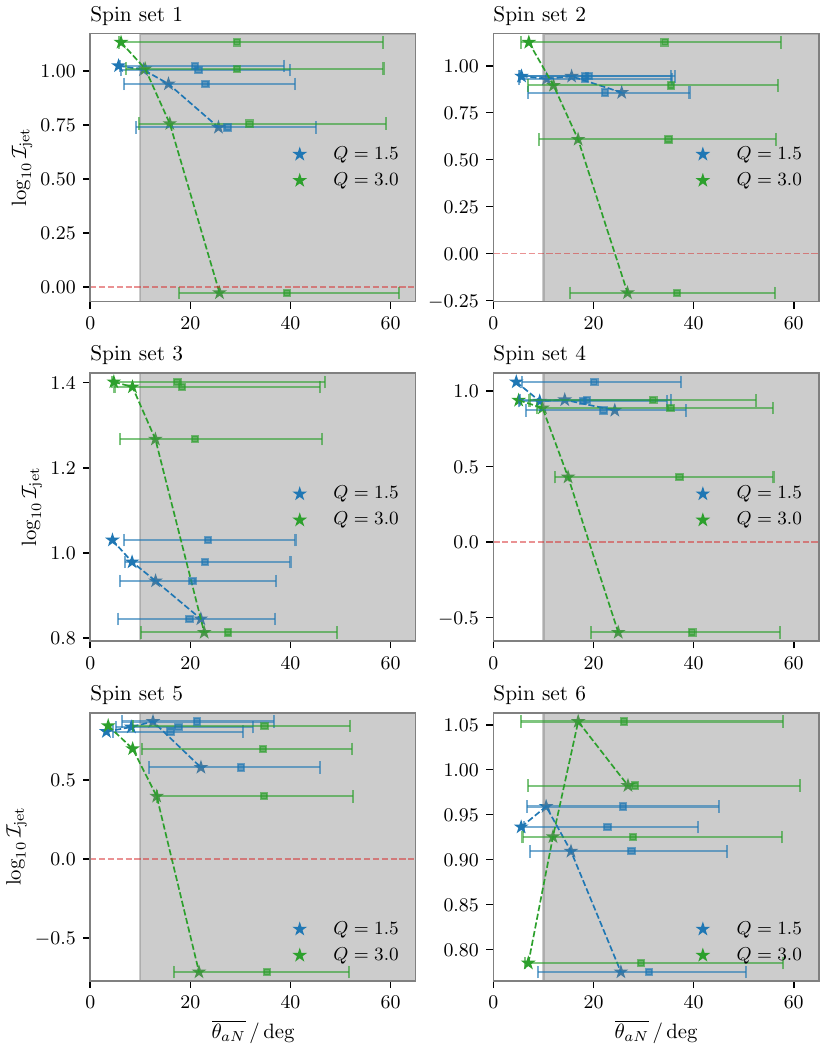}
    \includegraphics[width=.48 \linewidth]{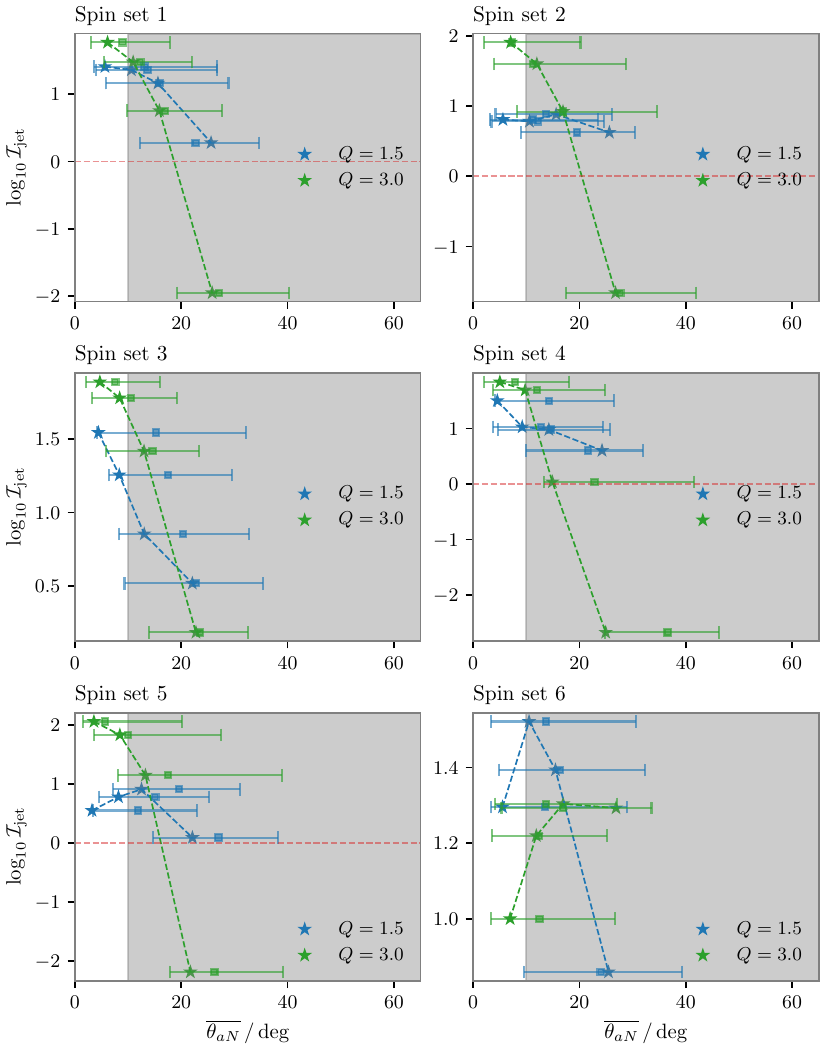}
    \caption{{\bf Overlap integrals results for SNR 15 and 30.} 
        Same as Fig.~\ref{fig:injection_overlap_extra} but for cases with SNRs of 15 (left) and 30 (right).
    }
    \label{fig:injection_overlap_extra}
\end{figure*}

Here we describe in more detail our results on injections with SNRs of 15 (similar to that of GW190521) and 30. Fig.~\ref{fig:injection_overlap_extra} shows the same as Fig.~\ref{fig:injection_overlap} in the main text, for these lower SNR cases.
First, we note the same trend discussed in the main text for ${\rm SNR}=50$ holds: the jet hypothesis is supported for small $\theta_{aN}^{\rm true}$ while the larger $\theta_{aN}^{\rm true}$ is, the more such hypothesis is rejected.
The main difference with the ${\rm SNR}=50$ case is that the corresponding evidences are of course less conclusive.
For ${\rm SNR}=15$, the jet hypothesis is supported with Bayes' factors up to $\simeq15$ in the best cases. This is, should the flare be really produced by a jet mechanism, any odds in favour of a true coincidence would be increased by a factor of up to 15. For the ${\rm SNR}=30$ case, these factor goes up to values of $100$. These results respectively imply that, on top of further confirming a true coincidence, the jet-nature of the flare would be moderately and strongly supported.
For jet-incompatible cases (with $\theta_{aN}^{\rm true} \in [10^\circ,30^\circ]$), we find that the jet hypothesis would only be strongly rejected for SNR$=30$, with a Bayes' factor of $1/100$. We have checked, nevertheless, that for larger opening angles, the jet hypothesis can be always completely ruled out.

\bibliography{bibliography.bib,LIGO_papers.bib,AGN_bibliography.bib}
\end{document}